\documentclass[aps,pre,twocolumn,groupedaddress,showpacs]{revtex4-1}

\usepackage{graphics,epsfig}
\usepackage{amsmath,amsfonts,mathdots,amssymb}

\begin{document}


\title{Phase-flip bifurcation and synchronous transition in unidirectionally coupled parametrically excited pendula}



\author{S. Satpathy}
\email{satya2047@gmail.com}
 \affiliation{Department of Physics \& Astronomy, National Institute of Technology Rourkela, India}
\author{B. Ganguli}%
 \email{biplabg@nitrkl.ac.in}
\affiliation{Department of Physics \& Astronomy, National Institute of Technology Rourkela, India}


\date{\today}

\begin{abstract}
Phase-flip bifurcation plays an important role in the transition to synchronization state in unidirectionally coupled parametrically excited pendula. In coupled identical system it is the cause of complete synchronization whereas in case of coupled non-identical system it causes desynchronization. In coupled identical systems negativity of conditional Lyapunov exponent is not always sufficient for complete synchronization. In complete synchronization state the largest conditional Lyapunov exponent and the second largest Lyapunov exponent are equal in magnitude and slope.
\end{abstract}

\maketitle

\section{Introduction}
The collective behavior of nonlinear dynamical systems is a fascinating area since the last three decades because of its inherent presence in nature and engineering systems. One of the most important of them is synchronization of chaotic systems. Chaos, due to its sensitive dependence on initial conditions, seems to defy synchronization. Since the wok of Pecora and Caroll \cite{Pecora1990}, synchronization of chaotic systems has been a subject of intense research.

 In the context of coupled chaotic systems, many different synchronization states are studied. They are: complete synchronization (CS) \cite{Fujisaka1983,Afraimovich1986,Pecora1990}, phase synchronization (PS) \cite{Rosenblum1996,Rosa1998}, lag synchronization (LS) \cite{Rosenblum1997}, generalized synchronization (GS) \cite{Rulkov1995,Abarbanel1996}, almost synchronization (AS) \cite{Afraimovich1986,Femat1999}, intermittent lag synchronization (ILS) \cite{Rosenblum1997,Boccaletti2000a}, imperfect phase synchronization (IPS) \cite{Zaks1999}, and intermittent generalized synchronization (IGS) \cite{Hramov2005a}.
 
This paper discussess synchronous and desynchronous transitions of unidirectionally coupled parametrically excited pendula due to phase-flip bifurcation. Parametrically excited pendulum shows three types of chaos namely tumbling, rotating, and oscillating depending upon the parameter settings \cite{Bishop1996}. We have taken oscillating chaos for our study \cite{Satpathy2017}. 

The phase-flip bifurcation was originally reported by Schuster and Wagner in a simple model of two delay-coupled phase oscillators \cite{Schuster1989}. Since then it is reported in varieties of time delay systems. It is observed in semiconductor lasers \cite{Kim2005}, delay coupled neuron models, ecological models, electronic circuits \cite{Prasad2008} in the periodic as well as chaotic regime, electrochemical cells \cite{Cruz2010}, a network of coupled bursting neurons \cite{Bhim2011}, relay coupled R\"{o}ssler oscillators \cite{Sharma2011}, dynamic environment coupling \cite{Sharma2012}, coupled Josephson junction neuron systems \cite{Seagall2014}, environmental diffusive coupling \cite{Sharma2016}, conjugate coupling with time delay \cite{Sharma2017}, experimental observation in brain \cite{Dotson2016} and in two inductively coupled glow discharge plasmas \cite{Chaubey2016}.

It is shown that phase-flip bifurcation is a fundamental transition in time-delay coupled, phase-synchronized systems \cite{Prasad2005,Prasad2006,Prasad2008}. When the time delay is varied, the synchrony between the oscillators undergoes a phase flip, i.e., the relative phase between the oscillators changes by $\pi$. This phenomenon is of broad relevance as it has been observed in regimes of amplitude death, periodic, quasiperiodic, and chaotic dynamics. It is recently investigated in detail by Prasad et al. \cite{Prasad2006,Prasad2008} in various types of delay-coupled oscillators, in which not only the phase but also the amplitude of the oscillators are affected by the coupling.

In the present work phase-flip bifurcation is observed where the coupling is simple diffusive. We show that it plays an important role in synchronization transition. In section \ref{IP} we discuss synchronization in coupled identical pendula while in section \ref{NP} the same is discussed in non-identical pendula. We summarize the results in the section \ref{C}.

\section{Coupled identical pendula}
\label{IP}

\begin{figure}[htp]
\includegraphics[scale=1]{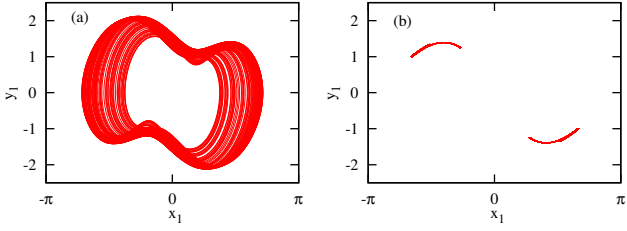}
\caption{(a) Phase portrait, (b) the points on the poincar\'e section, for $B = 1.0$, $\omega = 2.0$ and $A = 3.25$}
\label{FIG1}
\end{figure}

The equations of the master and slave  systems are given by
 \begin{eqnarray}
\dot{x}_1 &=& y_1 \nonumber \\ 
\dot{y}_1 &=& - By_1 - (1 + A\cos z)\sin x_1 ,\nonumber \\ 
\dot{x}_2 &=& y_2  \\ 
\dot{y}_2 &=& - By_2 - (1 + A\cos z)\sin x_2\nonumber +  k(x_1 - x_ 2), \\ \nonumber
\dot{z} &=& \omega, 
\label{CPEPDR}
\end{eqnarray}
\noindent where the subscript $1$ represents the master (drive) and  $2$ the slave (response). $B = 1.0$, $A = 3.25$, and $\omega = 2.0$, are the parameters chosen to produce chaotic behavior as shown in Fig. \ref{FIG1} \cite{Satpathy2017}. The initial conditions are taken as $x_1(0) = 0.3\pi$, $x_2(0)$, $y_1(0)$, $y_2(0)$ and $z(0)$ as zero.

\begin{figure}[h]
\centering
\includegraphics[scale=1.0]{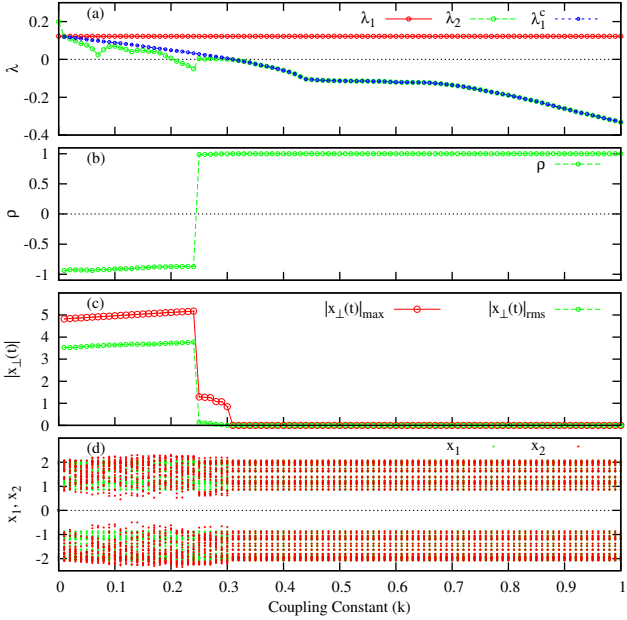}
\caption{The variation as a function of coupling constant of (a) The two largest Lyapunov exponents $(\lambda_1, \lambda_2 )$, and the largest conditional Lyapunov exponent $(\lambda^c_1)$, (b) Pearson correlation coefficient, (c) The average distance from the CS manifold $|x_\perp|_{rms}$ (green) and its maximum observed value $|x_\perp|_{max}$ (red), (d) Bifurcation diagram for $x_1$ and $x_2$ calculated from poincar\'e sections. Initial condition: $x_1(0) = 0.3\pi$, $x_2(0) = y_1(0) = y_2(0) = z (0) = 0$.}
\label{FIG2}
\end{figure}
At $k = 0$, the pendula have two positive, two negative and one zero Lyapunov exponents. That means the pendula are in hyper-chaotic state. Since the original system is non-autonomous, the zero exponent corresponds to the equation $\dot{z} = \omega$,  and are insensitive to the coupling. Fig. \ref{FIG2}(a) shows the largest Lyapunov exponent ($\lambda_1$) is positive for all values of coupling constant. The second largest Lyapunov exponent ($\lambda_2$) becomes negative at $k = 0.21$ and the systems become chaotic. There is a discontinuous transition in $\lambda_2$ at $k = 0.25$, signifying phase-flip bifurcation. The bifurcation is confirmed by linear cross-correlation coefficient.

The linear (Pearson) cross-correlation coefficient between two temporal series $x_1(t)$ and $x_2(t)$ is given by
\begin{equation}
\rho = \frac{\left\langle(x_1 - \langle x_1\rangle)(x_2 - \langle x_2\rangle)\right\rangle}{\sqrt{\left\langle (x_1 - \langle x_1\rangle)^2\right\rangle}\sqrt{\left\langle (x_2 - \langle x_2\rangle)^2\right\rangle}},
\end{equation}
where $\langle ... \rangle$ is the average over number of realization of the temporal series. The correlation coefficient takes the value between $-1$ and $1$. If it is zero the systems are uncorrelated and negative means the systems are negatively correlated \cite{Bove2004}. At the bifurcation point the Pearson correlation coefficient changes sign from negative to positive, shown in Fig. \ref{FIG2}(b). This confirms phase-flip bifurcation.

The largest conditional ($\lambda^c_1$) and $\lambda_2$ become negative for $k \geq 0.31$, which shows CS. To confirm this we calculate the average distance from CS manifold $|\mathbf x_\perp|_{rms}$ and its maximum value  $|x_\perp|_{max}$. $|\mathbf x_\perp|_{rms}$ for unidirectional coupling is defined as \cite{Gauthier1996}
\begin{equation}
|\mathbf x_\perp|_{rms} = \lim_{T \to \infty}\frac{1}{T - T0}\int_{T0}^T \mid {\mathbf x_1}(t) - {\mathbf x_2}(t) \mid dt,
\end{equation}
here $T$ is the total time of computation and $T0$ is the transient time. $|\mathbf x_\perp|_{rms}$ determines the global stability, while the maximum value of transverse distance from CS manifold  $|x_\perp|_{max}$ determines local stability of the CS state. It is clear from Fig. \ref{FIG2}(c) that $|x_\perp|_{max} \to 0$ for $k \geq 0.31$. These results confirm the systems are in CS for $k \geq 0.31$. This is also evident from the bifurcation diagram Fig. \ref{FIG2}(d).

\begin{figure}[htp]
\centering
\includegraphics[scale=1.0]{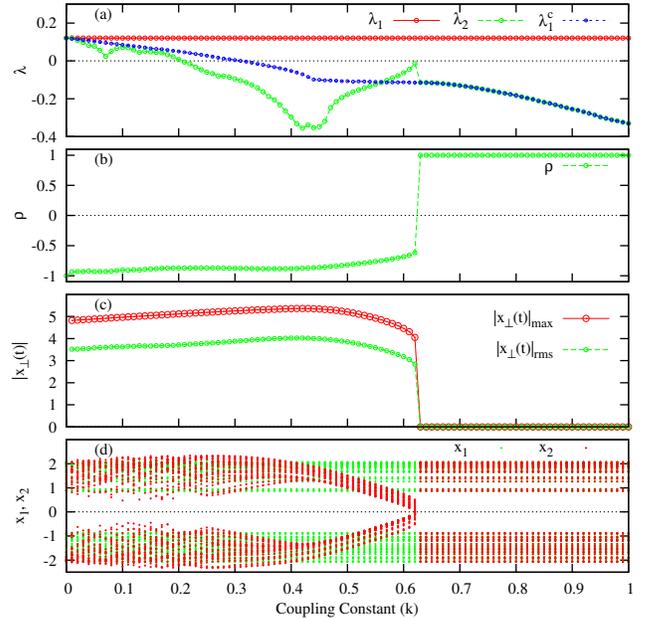}
\caption{Variations as a function of coupling constant of (a) The two largest Lyapunov exponents $(\lambda_1, \lambda_2 )$, and the largest conditional Lyapunov exponent $(\lambda^c_1)$, (b) Pearson correlation coefficient, (c) The average distance from the CS manifold $|x_\perp|_{rms}$ (green) and its maximum observed value $|x_\perp|_{max}$ (red), (d) Bifurcation diagram for $x_1$ and $x_2$ calculated from poincar\'e sections. Initial condition: $x_1(0) = 0.5\pi$, $x_2(0) = -0.5\pi$, $y_1(0) = y_2(0) = z(0) = 0$.}
\label{FIG3}
\end{figure}

The result, CS for $k \geq 0.31$ changes when different initial conditions are taken. We repeated our calculations for initial condition $x_1(0) = 0.5\pi$, $x_2(0) = -0.5\pi$, $y_1(0)$, $y_2(0)$ and $z(0)$ as zero. The variation of the two largest Lyapunov exponents ($\lambda_1$ and $\lambda_2$) and the largest conditional Lyapunov exponent $\lambda^c_1$ as a function of coupling constant are shown in Fig. \ref{FIG3}(a). The figure shows  $\lambda_2$ is negative for $k \geq 0.21$ and $\lambda^c_1$ is negative for $k \geq 0.31$.
The bifurcation diagram Fig. \ref{FIG3}(d)  shows the systems  do not synchronize at $k = 0.31$. The  discontinuity in $\lambda_2$ at $k = 0.63$ signifies the occurrence of phase-flip bifurcation. To confirm the occurrence of bifurcation we calculate Pearson correlation coefficient which is plotted in Fig. \ref{FIG3}(b). The figure shows the coefficient changes sign from negative to positive and attains the value $1$ at $k = 0.63$. This confirms the bifurcation to CS state at $k = 0.63$.
To further establish the result we calculate the transverse distance from CS manifold and its maximum value also and shown them in Fig. \ref{FIG3}(c). Both the values tend to zero for $k \geq 0.63$ and therefore the systems are in CS and the state is stable.

A careful observation of the Fig. \ref{FIG3}(a) shows when $\lambda_2$ and $\lambda_1^c$ overlap with equal slope, the systems are in CS. We find similar result for other initial conditions, for example Fig. \ref{FIG2}(a). It is possible that they may have same value but with different slopes at certain $k$. In that case, there is no CS at those values of $k$.

\begin{figure}[htp]
\centering
\includegraphics[scale=0.92]{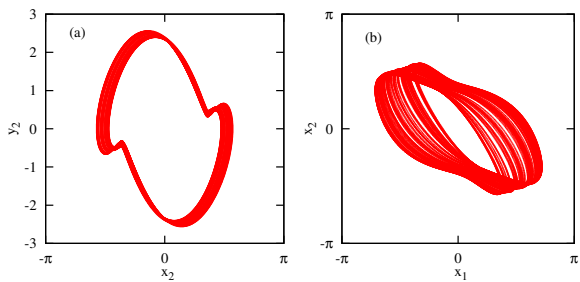}
\caption{Projections of the attractor of the coupled system on the plane $x_2-y_2$ (a), and on the plane $x_1-x_2$ (b), showing GS-like state for $k = 0.62$.}
\label{FIG4}
\end{figure}

The state is not CS for $0.31 \leq k \leq 0.62$ though $\lambda_1^c$ is negative in this range. Generally the negativity of $\lambda_1^c$  indicates the existence of stable synchronized manifold. But the bifurcation diagram Fig. \ref{FIG3}(d) shows the state is not stable with respect to variation in coupling constant. We get similar result in coupled non-identical system also. We call this synchronized state as ``GS-like''. This terminology is taken from our result found in non-identical systems discussed in section \ref{NP}. The phase portrait of the second system and the GS-like state are shown in Fig. \ref{FIG4}. From Fig. \ref{FIG3}(d) and Fig. \ref{FIG4}(a) we observe that the attractor size  increases suddenly at $0.63$ where phase-flip bifurcation takes place.

For initial condition, $x_1(0) = 0.3\pi$, $x_2(0) = y_1(0) = y_2(0) = z (0) = 0$, the phase-flip bifurcation occurs first as we change the coupling constant and $\lambda_1^c$  becomes negative later. The CS is achieved in this case at the value where $\lambda_1^c$ becomes negative. Whereas for initial condition, $x_1(0) = 0.5\pi$, $x_2(0) = -0.5\pi$, $y_1(0) = y_2(0) = z(0) = 0$, $\lambda_1^c$ becomes negative first and the phase-flip bifurcation occurs later. In this case CS is achieved at the bifurcation point instead when $\lambda_1^c$ becomes negative. This clearly shows, though negativity of $\lambda_1^c$ is the necessary condition, the phase-flip bifurcation ensures the CS state.

\begin{figure}[htp]
\centering
\includegraphics[scale=1]{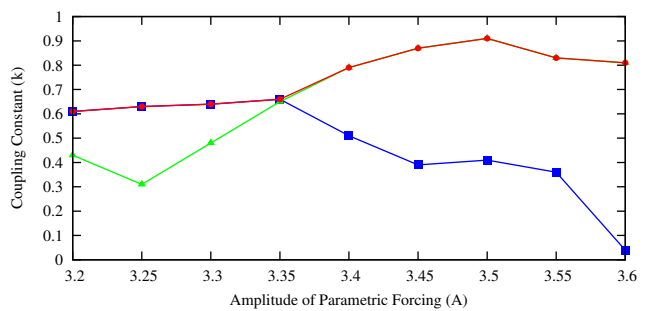}
\caption{The Bifurcation diagram for different values of amplitude of parametric forcing $A$. Onset of CS (red line with circle); negativity of largest conditional Lyapunov exponent (green line with triangle); The phase-flip bifurcation point (blue line with square).}
\label{FIG5}
\end{figure}

This result is further established by repeating the calculation for different values of the amplitude of parametric forcing ($A$), i.e., for different set of pendula. The calculations are summarized in Fig. \ref{FIG5}. The figure has two regimes. In the lower $A$ regime ($\leq 3.35$), $\lambda_1^c$ becomes negative first but the phase-flip bifurcation and onset of CS matches later. Whereas in the higher $A$ ($> 3.35$) regime phase-flip bifurcation occurs first but negativity of  $\lambda_1^c$  and onset of CS matches later. In the lower regime of $A$, CS is achieved for almost same value of coupling constant at $k = 0.6$. Since the bifurcation occurs after negativity of $\lambda_1^c$, therefore GS-like state is observed in the region. When $A$ is increased beyond $3.35$ up to $3.5$ the coupling constant at which CS occur increases and it decreases when $A> 3.5$. There is no GS-like state in this region because the bifurcation occurs before the negativity of $\lambda_1^c$.

\section{Coupled non-identical pendula}
\label{NP}
\begin{figure}[htp]
\includegraphics[scale=1]{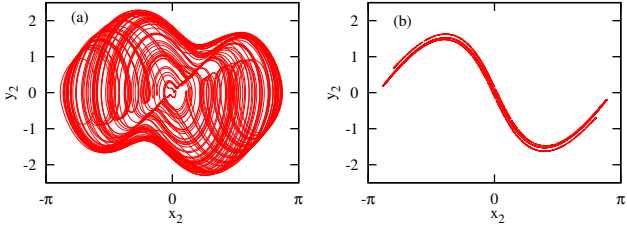}
\caption{(a) Phase portrait, (b) the points on the poincar\'e section, for $B = 1.0$, $\omega = 2.0$ and $A_2 = 3.55$.}
\label{FIG6}
\end{figure}

The equations of the master-slave systems are given by
\begin{eqnarray}\label{CNIPEPDR}
\dot{x}_ 1 &=& y_1, \nonumber \\ 
\dot{y}_1 &=& - By_1 - (1 + A_1\cos z)\sin x_1,\nonumber \\ 
\dot{x}_ 2 &=& y_2,  \\ 
\dot{y}_2 &=& - By_2 - (1 + A_2\cos z)\sin x_2  +  k(x_1 - x_2),\nonumber \\ 
\dot{z} &=& \omega, \nonumber
\end{eqnarray}
where the subscripts $1$ and $2$ refer to master (drive) and slave (response) respectively, and $k$ is the coupling constant.  $B = 1.0$, $A_1 = 3.25$, $A_2 = 3.55$ and $\omega = 2.0$, are the parameters chosen to produce chaotic behavior as shown in Fig. \ref{FIG1} and \ref{FIG6} \cite{Satpathy2017}.\\

\begin{figure}[t]
\centering
\includegraphics[scale=1]{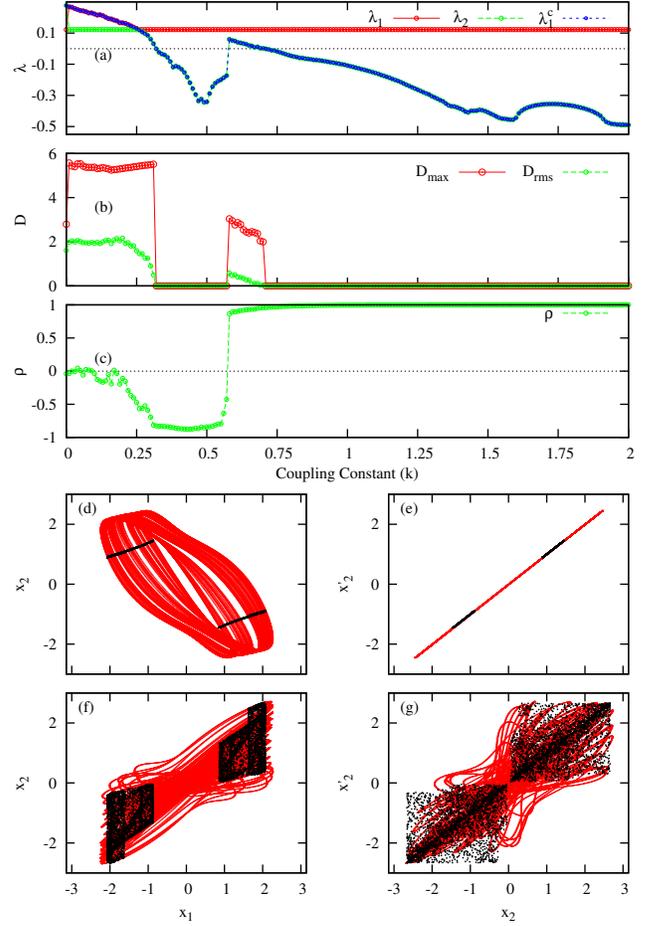}
\caption{(a)The two largest Lyapunov exponents ($\lambda_1$, $\lambda_2$) and the conditional Lyapunov exponent ($\lambda^c_1$) vs $k$, (b) The averaged distances $D$ of the response and its auxiliary vs $k$. They approach zero at different threshold couplings, indicating the achievement of generalized synchronizations, (c) Cross-corelation function as a function of vs $k$, (d, f) Projection of the phase trajectories of the coupled system \eqref{CNIPEPDR} in the space $(x_1, x_2)$, (e, g) Projections of the phase trajectories onto the plane of similar variables in the response and auxiliary systems $(x_2 , x_2')$, (d, e) $k = 0.57$, (f, g) $k = 0.58$, The black dots show the points on the Poincar\'e map.}
\label{FIG7}
\end{figure}

The Lyapunov spectrum have two positive, two negative, and one zero exponents at $k = 0$. The two largest Lyapunov exponents ($\lambda_1$, $\lambda_2$), and the largest conditional Lyapunov exponent ($\lambda^c_1$)  are plotted as a function of coupling constant, shown in Fig. \ref{FIG7}(a). $\lambda_2$ is equal to $\lambda^c_1$ for $k\geq 0.21$.  The variation of $\lambda^c_1$ shows two regions of synchronization. The first is for $0.32 \leq k \leq 0.57$ and second for $k \geq 0.71$, where $\lambda^c_1$ is negative.

We apply various methods to predict synchronized states and their stability. The first method is the auxiliary system approach \cite{Abarbanel1996}. It considers an auxiliary system identical to the response. When the drive system drives both the response and auxiliary systems, the vector fields in the respective phase spaces of the response and auxiliary are identical and the systems can evolve on identical attractors.

The equations of the auxiliary system in the present case are given by
\begin{eqnarray}
\dot{x}^\prime_ 2 &=& y^\prime_2,  \nonumber\\ 
\dot{y}^\prime_2 &=& - By^\prime_2 - (1 + A_2\cos z)\sin x^\prime_2  +  k(x_1 - x^\prime_2), \\ 
\dot{z} &=& \omega, \nonumber
\end{eqnarray}
The transverse distance of auxiliary system from the response system is calculated from 
\begin{equation}
D_{rms} = \lim_{t \to \infty}\frac{1}{T - T0}\int_{T0}^T \mid {\mathbf X_2}(t) - {\mathbf X^\prime_2}(t) \mid dt,
\end{equation}
where ${\mathbf X_2}$ and $\mathbf X^\prime_2$ are response and auxiliary systems respectively. When $D$ approaches zero, the state vectors of the response and auxiliary systems begin to coincide with each other. This indicates the presence of GS. $D_{rms}$ and its maximum observed value ($D_{max}$) are shown in Fig. \ref{FIG7}(b). Synchronization takes place for values of $k$ where both $D_{rms}$ and  $D_{max}$ are zero. There are two intervals of $k$ which correspond to two synchronization regimes. The same result is also observed from $\lambda^c_1$. These intervals are separated by a region ($0.58 \leq k \leq 0.7$) of asynchronous behavior where $\lambda^c_1$ is positive. It is also observed in synchronization of unidirectionally coupled electronic circuits {\cite{Rulkov2001,Rulkov2001b}}.

The chaotic attractors for $k=0.57$ are shown in Fig. \ref{FIG7}(d and e) and for $k = 0.58$ in Fig. \ref{FIG7}(f and g). It is clear from the two figures that GS terminates at $k=0.58$. It is due to the emergence of sporadic outbursts of non-identical behavior in the response and auxiliary systems. The reason for the outbursts is the transition of Pearson cross-correlation coefficient from negative to positive as shown in Fig. \ref{FIG7}(c). This transition indicates the occurrence of phase-flip bifurcation. This can also be seen in the discontinuous transition of $\lambda_2$ in Fig. \ref{FIG7}(a). The correlation function remains almost constant value in the regions of synchronization. The function, in the first synchronization interval, is near to $-1$. But it is almost equal to $1$ in the second synchronization region. Therefore the synchronized state is AS instead of GS  \cite{Satpathy2018}. We carried out further calculations to establish GS and AS region of synchronization. For this purpose we use method based on nearest neighbor. 

We select the nearest neighbor ${\mathbf u}^{nn}$ of ${\mathbf u}^n$ for $n = 1,.\,.\,.\, ,N$ and compute the average distance of the corresponding image points ${\mathbf v}^n$ and ${\mathbf v}^{nn}$ given by
\begin{equation}
d_{\mathbf{xy}} = \frac{1}{N\delta_\mathbf{y}}\sum_{n=1}^N \parallel{\mathbf v}^n-{\mathbf v}^{nn}\parallel.
\label{MD1}
\end{equation}
This mean distance is normalized by the average distance $\delta_\mathbf y$ of randomly chosen states of the response system \cite{Parlitz1996,Parlitz2012}. 

\begin{figure}[htp]
\centering
\includegraphics[scale=1.0]{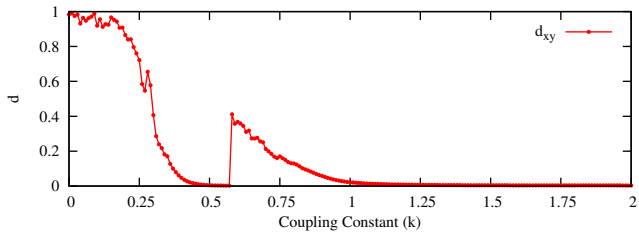}
\caption{The quantitative measure $d$ vs the coupling strength $k$.}
\label{FIG8}
\end{figure}

To calculate $d_{\mathbf{xy}}$ the variables $x_1$ and $x_2$ are considered for the attractor reconstruction \cite{BAbarbanel1996}. The data is sampled at time step $45\Delta t$. The time delays are taken to be $4$ calculated from mutual information. The embedding dimension for the drive is found out to be $5$. The same embedding dimension is chosen for the response. The Fig. \ref{FIG8} shows $d_{xy}$ tends to zero in the two regions of synchronization. This confirms the existence of GS in these regions.

There exist a variation of false nearest neighbor method \cite{Arnhold1999,Quiroga2000,Schmitz2000,Bhattacharya2003}. This method considers $k$ nearest neighbors of drive ($X$) and response ($Y$). Based on this concept three closely related indexes $S$, $H$, and $E$ are defined, depending on three different mean Euclidean distances $R_n$. $S$ is defined as
\begin{equation}
\label{SXY}
S^{(k)}(X|Y) = \frac{1}{N} \sum_{n = 1}^N\frac{R_n^{(k)}(X)}{R_n^{(k)}(X|Y)}.
\end{equation}
where $R_n^{(k)}(X)$ is the squared mean Euclidean distance to its $k$ nearest neighbors and $R_n^{(k)}(X|Y)$ is the conditional mean squared Euclidean distance. 

$H$ is defined as
\begin{equation}
H^{(k)}(X|Y) = \frac{1}{N}\sum_{n = 1}^N \frac{R_n(X)}{R_n^{(k)}(X|Y)}.
\end{equation}
where $R_n(X)$ is mean squared distance to random points.

$E$ is defined as
\begin{equation}
\label{EXY}
E^{(k)}(X|Y) = \frac{1}{N} \sum_{n = 1}^N\frac{R_n^{\prime(k)}(X)}{R_n^{(k)}(X|Y)}.
\end{equation}
where $R_n^{\prime(k)}(X)$ is the squared mean distance of $k+1$ to $2k$ nearest neighbors.

If $S^{(k)}(X|Y) \ll 1$, $X$ and $Y$ are independent or unsynchronized. $S^{(k)}(X|Y) \to 1$ indicates the occurrence of  GS. $H^{(k)}(X|Y)$ is zero if $X$ and $Y$ are completely independent, while it is positive if nearness in $Y$ also implies nearness in $X$. It would be negative if close pairs in $Y$ correspond mainly to distant pairs in $X$. This is very unlikely but not impossible. Since $E$ is not normalized, it should be treated as a regulatory statistic or comparative measure without highlighting the absolute values.

\begin{figure}[htp]
\centering
\includegraphics[scale=1]{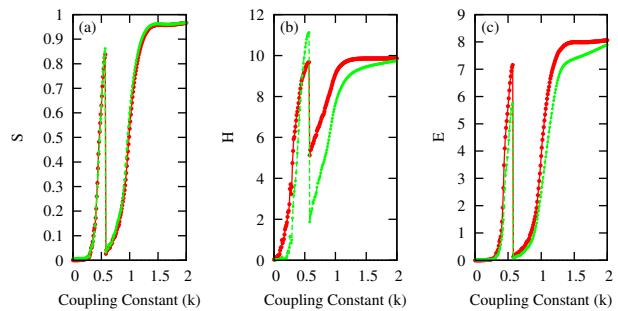}
\caption{The indexes plotted as a function of coupling strength. The $(X|Y)$ dependencies are plotted in red and $(Y|X)$ in green.}
\label{FIG9}
\end{figure}

\begin{figure}[htp]
\centering
\includegraphics[scale=1]{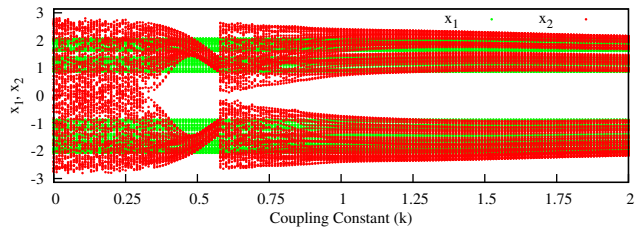}
\caption{Bifurcation digram for the systems Eq. \eqref{CNIPEPDR}.}
\label{FIG10}
\end{figure}

The indexes $S$, $H$, and $E$ are calculated  from the already reconstructed attractors. $S(X|Y)$, $S(Y|X)$, $H(X|Y)$, $H(Y|X)$, $E(X|Y)$, and $E(Y|X)$ for $10$ nearest neighbors are plotted as a function of coupling constant, shown in Fig. \ref{FIG9}. The figure shows that the indexes do not remain constant in the first synchronization region. This means the synchronization state is not stable with respect to coupling constant. It is also evident from the bifurcation diagram, Fig. \ref{FIG10}. Since this state is predicted to be GS in previous two methods while the third method contradicts, therefore we call the synchronized state as ``GS-like''. 

The indexes remain almost constant in the second region of synchronization. Both the indexes $E$ tell us which system is drive and which one is response. The figure shows $E(X|Y) > E(Y|X)$, therefore the response system is more active than the drive. In case of drive-response relation the drive represents the dynamics of drive alone; but the response represents the dynamics of the drive and the response both. So the response has more information about the drive.

\begin{figure}[htp]
\centering
\includegraphics[scale=1]{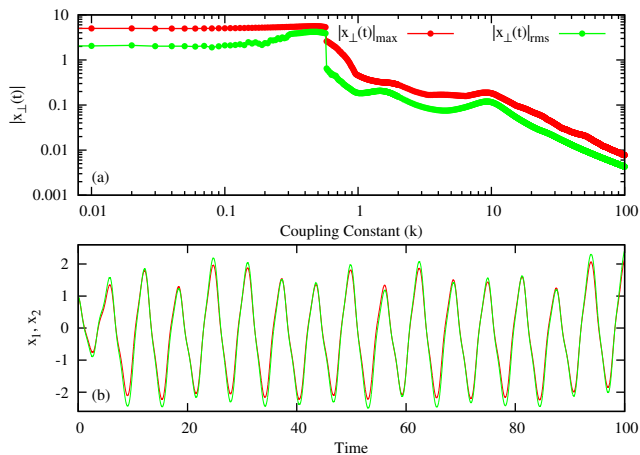}
\caption{(a) The average distance from the CS manifold $|x_\perp(t)|_{rms}$ and its maximum value  $|x_\perp(t)|_{max}$ as a function  of coupling strength $k$ for the systems Eq. \eqref{CNIPEPDR}, (b) Time series of the two systems showing AS.}
\label{FIG11}
\end{figure}

To prove AS in the second synchronization region, we calculate the transverse distance from the CS manifold $|x_\perp|_{rms}$ and its maximum value $|x_\perp|_{max}$ as shown in Fig. \ref{FIG11}(a). It is clear that these values decrease with increase in coupling strength. This indicates the synchronized state is AS.  The time series plot, Fig. \ref{FIG11}(b), of the variables $x_1$ and $x_2$  shows complete phase matching but the amplitudes are slightly different. The results can be seen in the bifurcation digram Fig. \ref{FIG10}. 

\begin{figure}[htp]
\centering
\includegraphics[scale=1]{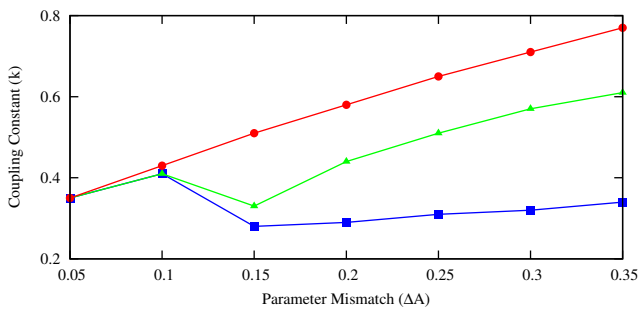}
\caption{The Bifurcation diagram for different values of $\Delta A$. Onset of AS is hown in red line (circles);  desynchronization regime is between green line (triangles) and red line (circles); GS-like state is between blue line (squares) and green line (triangles).}
\label{FIG12}
\end{figure}

The coupling constant $k = 0.71$ is the transition point from asynchronous state to AS state. The calculations are repeated for different set of systems where the amplitude of parametric forcing of the second system is varied. We summarize the results in Fig. \ref{FIG12}. The coupling constant for transition to AS increases linearly with increase in mismatch of parametric forcing ($\Delta A$) as shown by red line. There is no GS-like state for low  $\Delta A$ ($\leq 0.1$) and the asynchronous state occupies very small region. The region of GS-like state increases marginally between $0.1-0.15$ of $\Delta A$ and asynchronous state increases drastically. The region of GS-like state increases significantly beyond $\Delta A = 0.15$ and the asynchronous state remains almost constant.

\section{Conclusion}
\label{C}
Generally the negativity of largest conditional Lyapunov exponent ($\lambda_1^c$) implies CS. This is not always true in the case of unidirectionally coupled identical parametrically excited pendula. Phase-flip bifurcation always accompanies synchronization transition. In some cases of initial conditions, phase-flip bifurcation occurs later than negativity of $\lambda_1^c$ when coupling constant is increased and then the bifurcation ensures CS. In such cases there exist a synchronization manifold between the occurrence of negativity of $\lambda_1^c$ and phase-flip bifurcation. The manifold is unstable with respect to coupling constant. We call this unstable synchronization as GS-like. In CS the second largest Lyapunov exponent ($\lambda_2$) and $\lambda_1^c$ are equal and both have same slope at that coupling constant.

In case of coupled non-identical pendula, there are two intervals of coupling constant which correspond to two synchronization regimes. The transition from the first synchronized state to asynchronous state is through phase-flip bifurcation. The synchronization manifold in the first synchronization regime is unstable with respect to coupling constant. Therefore the synchronized state is GS-like. The second synchronization regime is AS. The coupling constant for transition to AS increases linearly with increase in mismatch of parametric forcing ($\Delta A$). There are no GS-like state for low  $\Delta A$ ($\leq 0.1$) and the asynchronous state occupies very small region. The region of GS-like state increases marginally between $0.1-0.15$ of $\Delta A$ and asynchronous state increases drastically. The region of GS-like state increases significantly beyond $\Delta A = 0.15$ and the asynchronous state remains almost constant.

\bibliography{Bibliography}

\end{document}